\documentclass[conference]{IEEEtran}
\IEEEoverridecommandlockouts

\usepackage[utf8]{inputenc}
\usepackage[T1]{fontenc}
\usepackage{cite}
\usepackage{amsmath,amssymb,amsfonts}
\usepackage{graphicx}
\usepackage{textcomp}
\usepackage{xcolor}
\usepackage{booktabs}
\usepackage{multirow}
\usepackage{array}
\usepackage{algorithm}
\usepackage{algpseudocode}
\graphicspath{{figures/}}

\usepackage[colorlinks=true,urlcolor=blue,linkcolor=black,citecolor=black]{hyperref}
\usepackage{eso-pic}

\let\OLDthebibliography\thebibliography
\renewcommand{\thebibliography}[1]{%
  \OLDthebibliography{#1}%
  \setlength{\itemsep}{2pt}%
  \setlength{\parskip}{0pt}%
}

\begin{document}

\title{An Uncertainty-Driven Hybrid Deep Learning Approach for\\
Broad-Coverage RF Modulation Recognition}

\author{
\IEEEauthorblockN{Nurettin Safak, Durdu Can Yerdeyatar, Muhammet Sefa Demirel, Alperen Marasli, Taha Eren Atmaca, Ozgun Ersoy}
\IEEEauthorblockA{Department of Electrical and Electronics Engineering, Ankara Yildirim Beyazit University, Ankara, Turkey}
\IEEEauthorblockA{nsafaked@gmail.com, yrdeytr4601@gmail.com, sefademirel56@gmail.com,\\
alperen.eem@gmail.com, tahaeren1883@gmail.com, ozgun.ersoy@aybu.edu.tr}
}

\AddToShipoutPictureFG*{%
  \ifnum\value{page}=1
    \AtPageLowerLeft{%
      \hspace{0.55in}\raisebox{0.55in}{%
        \footnotesize\textbf{979-8-3195-4709-5/26/\$31.00~\textcopyright{}2026 IEEE}%
      }%
    }%
  \fi
}

\maketitle

\begin{abstract}
Automatic RF modulation recognition is of critical importance in spectrum monitoring, electronic warfare, and cognitive radio applications, where low signal-to-noise ratio (SNR) conditions and the growing diversity of modulation schemes limit the performance of existing methods. This paper proposes an uncertainty-driven hybrid deep learning architecture for recognizing RF signals over a broad modulation space. The proposed approach carries out a multi-stage classification process by combining spectral information obtained through low-cost FFT-based preprocessing with time-frequency features extracted from short-time Fourier transform (STFT) spectrograms. The architecture comprises a 2D convolutional neural network (2D CNN)-based path for fast, low-latency primary classification, MC Dropout-supported Bayesian uncertainty estimation for assessing classification reliability, and a BiLSTM-based secondary decision mechanism activated under high-uncertainty conditions. The proposed system is evaluated in a controlled simulation environment spanning different SNR levels and modulation classes. Experimental results show that the primary 2D CNN path achieves 83.3$\pm$0.7\% accuracy with an inference time of only 0.138\,ms per sample, providing superior performance compared with traditional rule-based and classical machine-learning approaches. Furthermore, the obtained findings reveal the limitations of compact spectral feature representations and classifiers lacking temporal modeling, particularly in disambiguating FSK-based modulations. The uncertainty estimation module offers promising results for detecting low-confidence decisions, and the proposed approach demonstrates the potential of a low-latency and scalable solution for real-time RF modulation recognition.
\end{abstract}


\IEEEpeerreviewmaketitle

\section{Introduction}

The electromagnetic spectrum is a finite, increasingly congested resource shared by modern communication, spectrum management, and electronic intelligence systems; growing device counts and signal-environment complexity make automatic spectrum monitoring and real-time transmission recognition a critical requirement~\cite{dobre2007}. At the heart of this task lies automatic modulation classification (AMC): determining a received signal's modulation type without prior information.

Traditional AMC relies on expert-knowledge-based rule sets such as likelihood-ratio tests and statistical moment features~\cite{dobre2007}, which fail noticeably under unknown carrier phase offset, expanding modulation pools, low-SNR conditions, and low-power-spectral-density spread-spectrum signals (DSSS, direct-sequence spread spectrum; FHSS, frequency-hopping spread spectrum). These limitations have markedly increased interest in data-driven deep-learning-based solutions over the past decade~\cite{oshea2017,west2017,oshea2018,peng2022,zhang2022dlamr}.

Neural networks trained end-to-end on raw IQ data or time-frequency representations eliminate the need for hand-crafted feature engineering and exhibit strong generalization~\cite{rajendran2018}. In particular, hybrid CNN--RNN architectures jointly model spatial and temporal features~\cite{xu2020,liao2021}. More recent work advances the field through robust global feature-extraction variants, state-space architectures, and broad surveys~\cite{qu2024global,zhang2024mamca,jafarigol2025aiml,padhya2025cnnlstm}. Nevertheless, existing studies remain deficient in three respects. First, most produce only a point estimate without an uncertainty measure, leading to overconfident misclassifications at low SNR and on out-of-distribution signals; uncertainty-based AMC has recently emerged~\cite{yang2025}, yet its integration with real-time hybrid routing remains under-addressed. Second, classes such as 4-FSK, DSSS, and FHSS are often excluded from evaluation, so reported accuracies understate the operational difficulty. Third, single-architecture solutions cannot dynamically manage the speed--accuracy trade-off.

To address these deficiencies, the proposed system combines, under a single decision flow, a fast FFT preprocessing layer, a real-time 2D CNN primary path, an MC Dropout-based Bayesian MLP uncertainty estimator, and a high-accuracy BiLSTM secondary path activated when uncertainty exceeds a predefined threshold -- dynamically managing the accuracy-latency trade-off while quantifying decision reliability.

The main contributions of this study are summarized as follows:
\begin{itemize}
\item An uncertainty-driven, multi-stage hybrid pipeline combining FFT-based fast parameter extraction ($O(N\log N)$, negligible overhead) with real-time 2D CNN classification (0.138\,ms),
\item A comparative analysis of rule-based, classical ML, and deep-learning approaches across modulation types and SNR levels,
\item Uncertainty-aware classification with MC Dropout and threshold-based dynamic routing to the BiLSTM secondary path,
\item The 3\% recall finding for 4-FSK, quantifying the limitation of FSK/4-FSK order disambiguation in compact spectral feature representations.
\end{itemize}

Field validation on real SDR hardware is left to future work; results here are from controlled simulation.

\section{System Model}

\subsection{Simulation Environment and Dataset}

A controlled simulation framework (NumPy/SciPy for signal generation and STFT, Python~3.12; models in PyTorch~2.12.0, trained on a single NVIDIA RTX~5070 laptop GPU) was used to ensure full control of SNR/channel conditions and reproducibility.

The dataset comprises 14 modulation classes in the three categories of Table~\ref{tab:classes}: $f_s=200$\,kHz, $f_\mathrm{sym}=10$\,kHz, signal length 4096 samples ($\approx$20.5\,ms); DSSS uses a 31-chip PN sequence, FHSS an 8-symbol hop; each sample has an independent random seed to prevent leakage. With 2040 samples/class over seven SNR levels ($-5$ to $+25$\,dB), an AWGN dataset of 199{,}920 samples is obtained, unit-power normalized and converted to $32\!\times\!32$ log-power spectrograms, with a 70\%/10\%/20\% (train/val/test) split.

\begin{table}[!h]
\caption{Taxonomy of 14 Modulation Classes}
\label{tab:classes}
\centering
\small
\begin{tabular}{ll}
\toprule
\textbf{Category} & \textbf{Classes} \\
\midrule
Analog          & AM, FM, NFM, WFM \\
Digital         & OOK, FSK, 4-FSK, BPSK, QPSK, \\
                & 8PSK, 16QAM, 64QAM \\
Spread Spectrum & DSSS, FHSS \\
\bottomrule
\end{tabular}
\end{table}

\subsection{STFT Representation}

IQ samples are transferred to the time-frequency plane via STFT ($N_\mathrm{fft}=64$, overlap $h=32$) and resized to a $32\!\times\!32$ log-power spectrogram for 2D CNN input -- a compact representation lowering memory/compute cost relative to higher-resolution spectrograms~\cite{zeng2019} and facilitating real-time operation~\cite{wang2020lightamc} (ablations in Section~\ref{sec:ablation}).

\section{Hybrid Model}

\subsection{Hybrid Pipeline and Decision Flow}

\begin{figure}[!t]
\centering
\includegraphics[width=0.99\columnwidth]{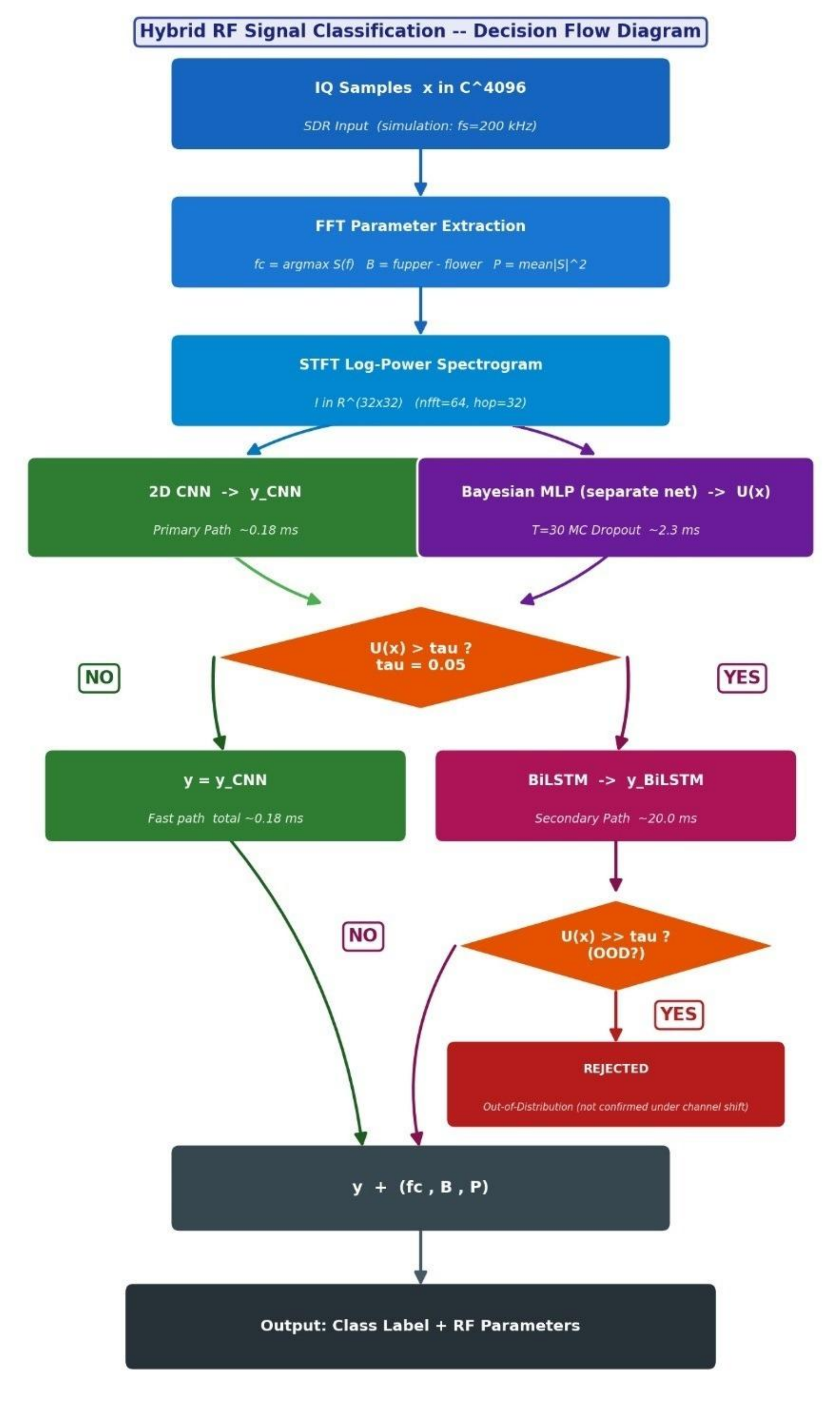}
\caption{Hybrid decision flow: FFT extraction and STFT, 2D CNN (0.138\,ms), Bayesian MLP uncertainty (0.7\,ms), and fusion --- BiLSTM (15.2\,ms) or OOD rejection at high uncertainty.}
\label{fig:flowchart}
\end{figure}

The proposed system has a four-stage decision flow shown in Fig.~\ref{fig:flowchart}. In \textit{Stage~1} (Parameter Extraction), the carrier frequency $f_c$, bandwidth $B$, and power $P$ are computed from the complex IQ stream as defined in Eq.~\eqref{eq:param}:
\begin{equation}\label{eq:param}
f_c = \arg\max_f S(f), \quad B = f_{\text{upper}} - f_{\text{lower}},\; S(f) \geq \theta S_{\max},
\end{equation}
where $\theta=0.5$ ($-3$\,dB threshold); this stage is $O(N\log N)$ and negligible compared with the 2D CNN path. \textit{Stage~2} produces a fast decision $\hat{y}_\mathrm{CNN}$ from the STFT spectrogram via the 2D CNN (0.138\,ms). \textit{Stage~3} computes an uncertainty score $\mathcal{U}(x)$ with the MC Dropout-based Bayesian MLP (0.7\,ms). \textit{Stage~4} compares this score with a threshold $\tau$: below it, the fast CNN decision is adopted directly; above it, the decision is delegated to the BiLSTM secondary path (15.2\,ms), which fully processes the temporal structure, and far above it the signal is rejected as out-of-distribution~\cite{hendrycks2017,shebert2023,zhang2023}. This layered design preserves real-time speed for most signals while engaging the costly secondary path only in difficult cases (Algorithm~\ref{alg:hybrid}).

\begin{algorithm}[!t]
\caption{Uncertainty-Driven Hybrid Classification}
\label{alg:hybrid}
\begin{algorithmic}[1]
\Require $\mathbf{x} \in \mathbb{C}^{4096}$: IQ samples
\Ensure $\hat{y}$: class label;\enspace $(f_c, B, P)$: RF parameters
\State $S(f) \gets \mathrm{FFT}(\mathbf{x})$
\State $f_c \gets \arg\max_f S(f)$;\enspace $B \gets f_{\mathrm{upper}}-f_{\mathrm{lower}}$;\enspace $P \gets \tfrac{1}{N}\textstyle\sum|S(f)|^2$
\State $\mathbf{I} \gets \mathrm{STFT\text{-}logpower}(\mathbf{x})$
\State $\hat{y}_{\mathrm{CNN}} \gets \mathrm{CNN}(\mathbf{I})$
\For{$t \gets 1$ \textbf{to} $T=30$}
    \State $\mathbf{p}^{(t)} \gets \mathrm{Softmax}(\mathrm{MLP}_{\mathrm{dropout}}(\mathbf{I}))$
\EndFor
\State $\mathcal{U}(\mathbf{x}) \gets \max_c\;\mathrm{std}\!\left[\{p_c^{(t)}\}_{t=1}^{T}\right]$
\If{$\mathcal{U}(\mathbf{x}) > \tau$}
    \State $\hat{y} \gets \mathrm{BiLSTM}(\mathbf{x})$
    \If{$\mathcal{U}(\mathbf{x}) \gg \tau$}
        \State \Return \textsc{Out-of-Distribution — Rejected}
    \EndIf
\Else
    \State $\hat{y} \gets \hat{y}_{\mathrm{CNN}}$
\EndIf
\State \Return $\hat{y},\;(f_c, B, P)$
\end{algorithmic}
\end{algorithm}

Table~\ref{tab:arch} compares the hybrid system with single approaches across five features: no component provides all of them, whereas the hybrid combines them.

\begin{table}[!h]
\caption{Architectural Feature Comparison}
\label{tab:arch}
\centering
\scriptsize
\setlength{\tabcolsep}{3.5pt}
\begin{tabular}{lccccc}
\toprule
\textbf{Method} & \textbf{Temp.} & \textbf{Unc.} & \textbf{RT} & \textbf{OOD} & \textbf{4-FSK} \\
\midrule
IQ Rule-Based    & ---        & ---        & \checkmark  & ---         & Low  \\
LinearSVC        & ---        & ---        & \checkmark  & ---         & Med. \\
XGBoost          & ---        & ---        & \checkmark  & ---         & Med. \\
Bayesian MLP     & ---        & \checkmark & \checkmark  & \checkmark  & Low  \\
2D CNN           & Part.      & ---        & \checkmark  & ---         & Med. \\
BiLSTM           & \checkmark & ---        & ---         & ---         & High \\
\textbf{Hybrid (Proposed)} & \checkmark & \checkmark & \checkmark & \checkmark & High \\
\bottomrule
\end{tabular}
\par\vspace{2pt}
\parbox{0.95\columnwidth}{\scriptsize Temp.: Temporal modeling; Unc.: Uncertainty estimation; RT: Real-time ($<$1\,ms); OOD: Out-of-distribution detection; Low/Med./High: 4-FSK recall level.}
\end{table}

\subsection{2D CNN Primary Path}

\begin{figure}[!t]
\centering
\includegraphics[width=0.99\columnwidth]{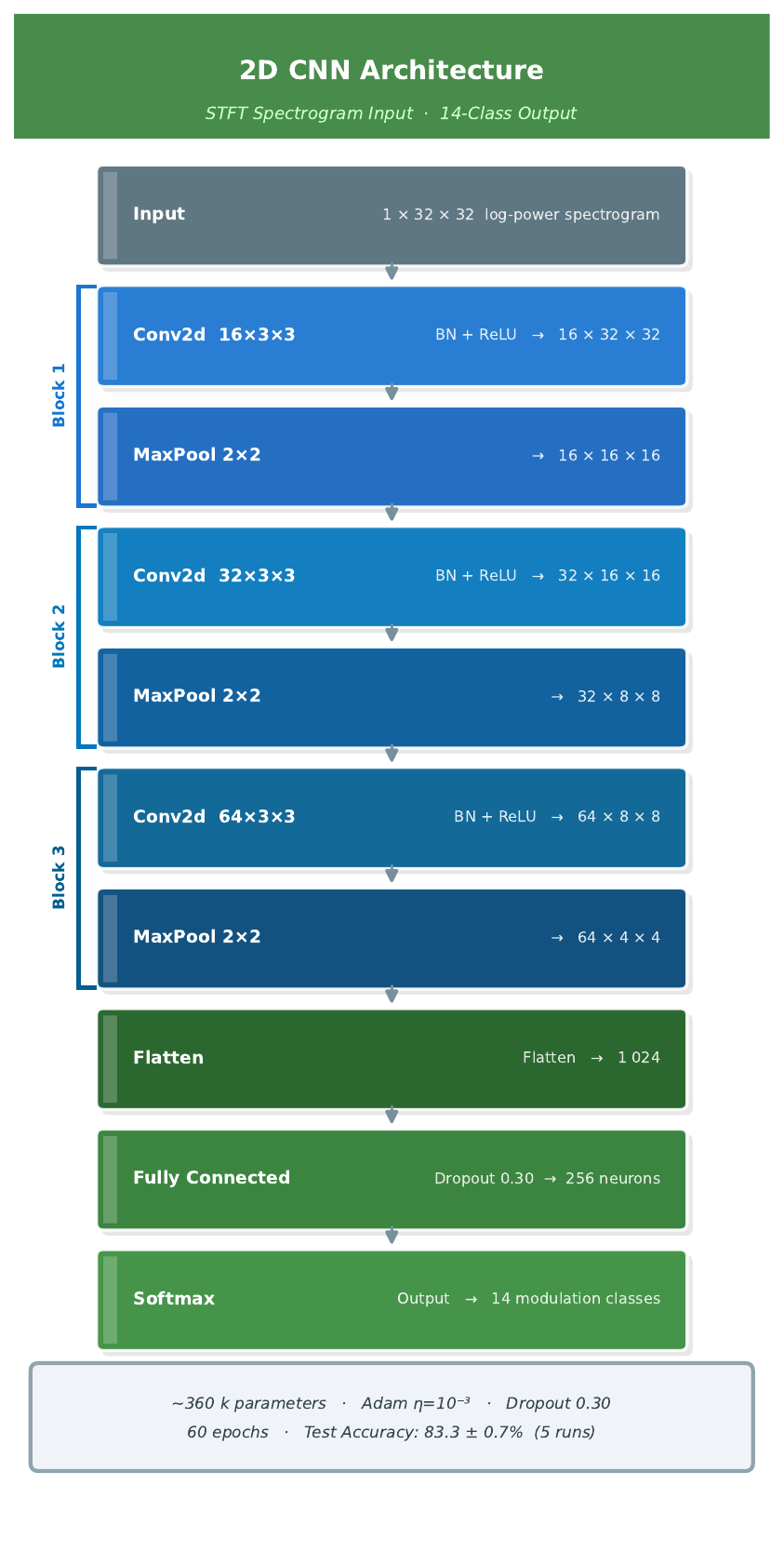}
\caption{2D CNN layer architecture: $1\!\times\!32\!\times\!32$ input, three convolutional blocks, 14-class output.}
\label{fig:cnnarch}
\end{figure}

The 2D CNN primary path (Fig.~\ref{fig:cnnarch}) has three convolutional blocks (16-, 32-, and 64-channel $3\!\times\!3$ convolutions with batch normalization, ReLU, and $2\!\times\!2$ max pooling), followed by a 256-neuron fully connected layer (40\% dropout) and a 14-class softmax output ($\approx$289{,}745 parameters). Trained with Adam ($\eta=10^{-3}$), cosine learning-rate scheduling, and 60 epochs using a weighted cross-entropy loss for class imbalance, its compact structure enables real-time operation at only 0.138\,ms per sample.

\subsection{Uncertainty Estimation and BiLSTM Secondary Path}

Uncertainty estimation builds on MC Dropout~\cite{gal2016,kendall2017}, which approximates Bayesian inference in deep Gaussian processes by keeping dropout active at inference; deep ensembles offer an alternative~\cite{lakshminarayanan2017}, but MC Dropout was preferred here for its single-model real-time advantage. The Bayesian MLP ($p=0.3$, three 256-neuron hidden layers) is run with $T=30$ stochastic passes; from the resulting probability vectors $\{\mathbf{p}^{(t)}\}_{t=1}^{T}$, uncertainty is the standard deviation over the highest-probability class:
\begin{equation}
\mathcal{U}(x) = \max_c\, \mathrm{std}\!\left[\{p_c^{(t)}\}_{t=1}^{T}\right].
\end{equation}
A low $\mathcal{U}(x)$ indicates model stability, a high value boundary cases or OOD signals. Under $\mathcal{U}(x) > \tau$ ($\tau=0.15$, empirically selected), the decision is routed to the BiLSTM secondary path.

The BiLSTM secondary path processes the first 1024 IQ samples (instantaneous envelope, frequency, and phase) with bidirectional LSTM layers (hidden size 256, 2 layers; 10\% validation, patience\,=\,10), implicitly encoding phase-transition patterns for high accuracy on classes requiring temporal structure such as FSK/4-FSK -- letting the system dynamically trade accuracy for latency as needed.

\section{Experimental Results}

\subsection{Resolution, FFT Length, and Epoch Ablation}
\label{sec:ablation}

\begin{figure}[!t]
\centering
\includegraphics[width=0.99\columnwidth]{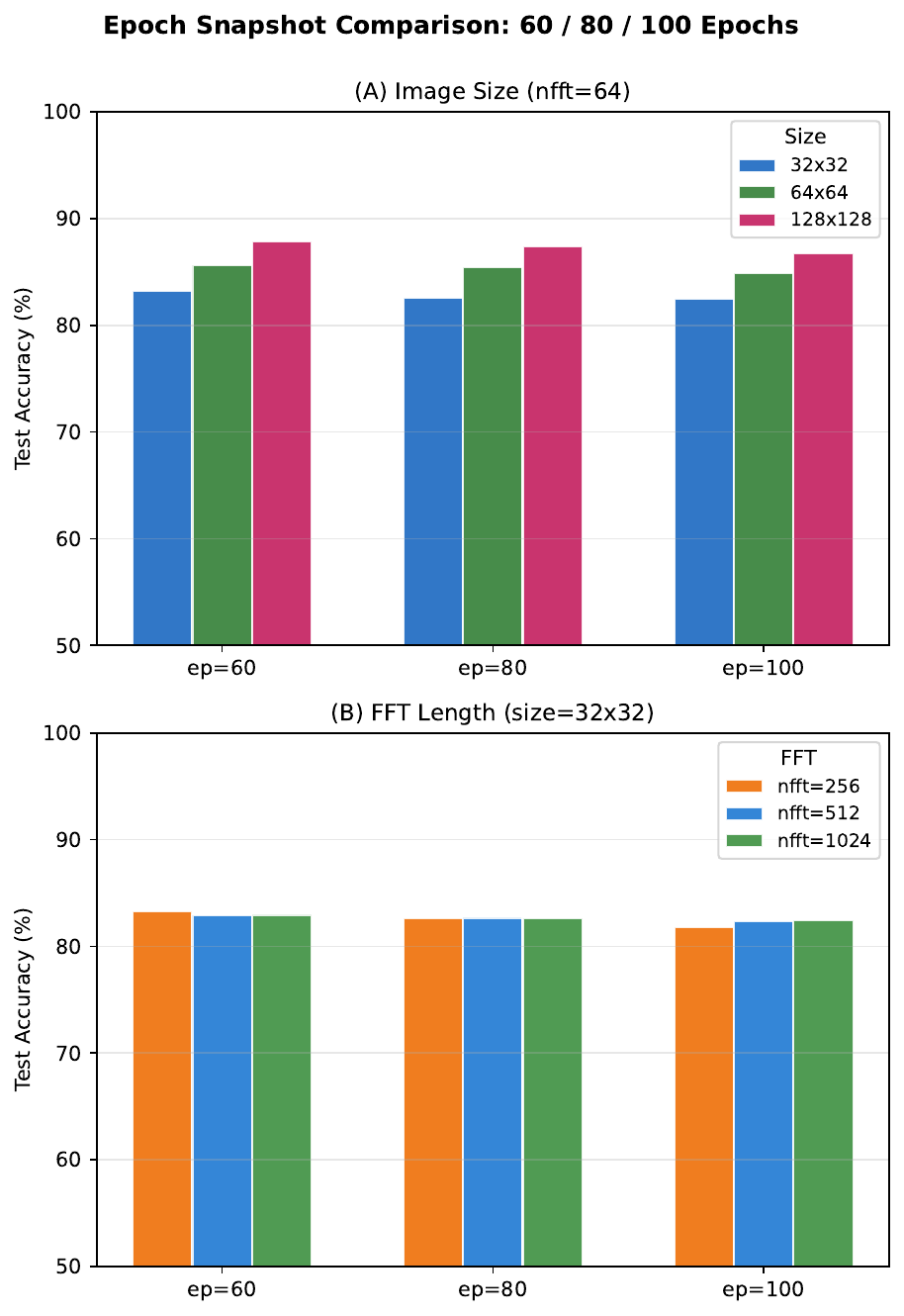}
\caption{Resolution and FFT length ablation (60/80/100 epochs). (A) image size ($N_\mathrm{fft}=64$); (B) FFT length ($32\!\times\!32$ image); FFT length has negligible effect.}
\label{fig:ablation}
\end{figure}

Two ablation experiments examined input resolution and FFT window length (Fig.~\ref{fig:ablation}; 60/80/100 epoch snapshots). With $N_\mathrm{fft}=64$ fixed, accuracy grows with image size (83.2\% at $32\!\times\!32$, 87.8\% at $128\!\times\!128$), but $128\!\times\!128$ raises the parameter count $\sim$14-fold for only a few points; with the image fixed at $32\!\times\!32$, varying $N_\mathrm{fft}\in\{256,512,1024\}$ leaves accuracy almost unchanged (82.5--83.3\%), showing FFT length has no meaningful effect once rescaled to a fixed size. Both experiments plateau around epoch 60 (100 epochs yield no gain and cause mild overfitting), so a $32\!\times\!32$ input and 60 epochs were chosen for memory-constrained real-time deployment.

\subsection{Comparative Performance by SNR}

\begin{figure}[!t]
\centering
\includegraphics[width=0.99\columnwidth]{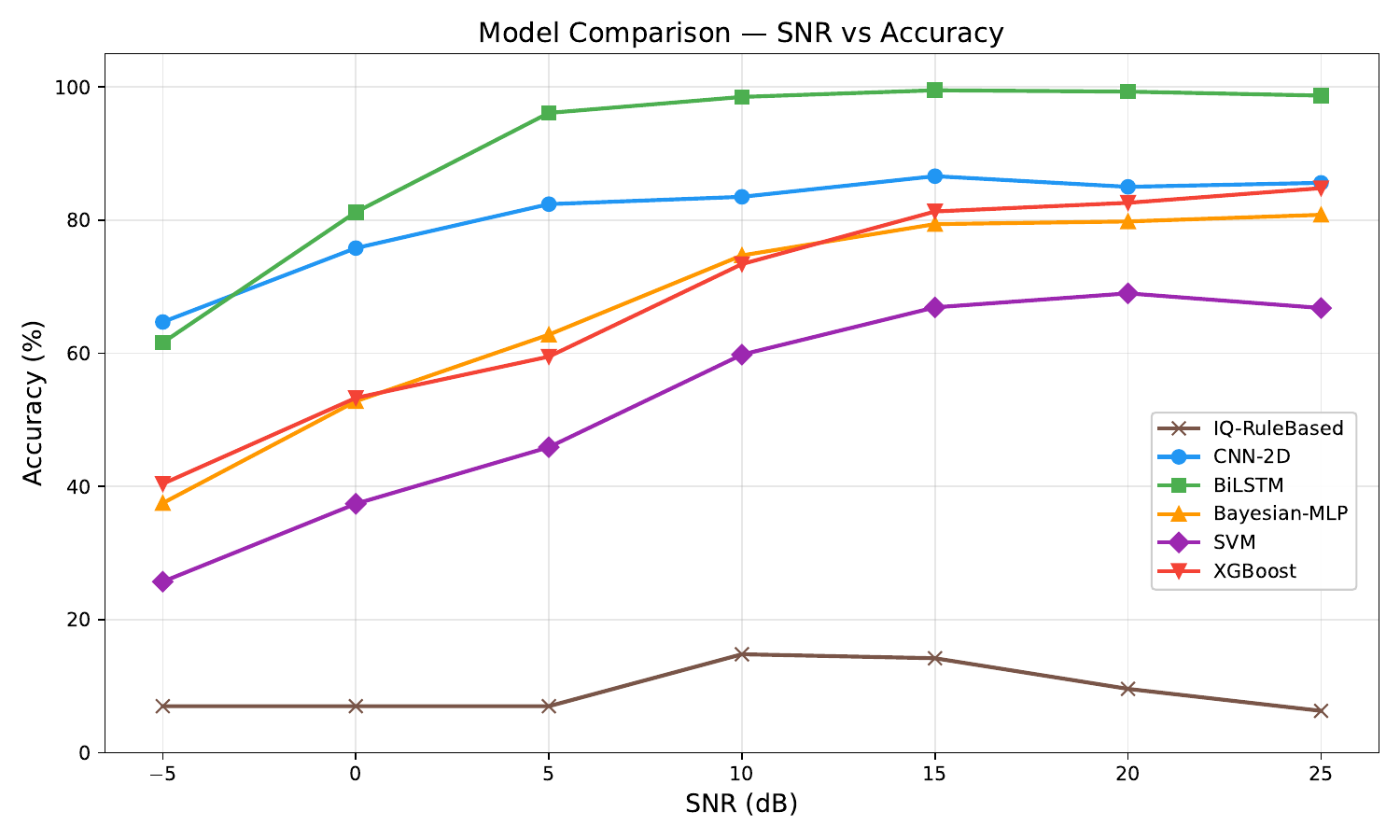}
\caption{Model accuracy comparison by SNR (14 classes, $-5$ to $+25$\,dB; companion benchmark, see Section~IV-B).}
\label{fig:snr}
\end{figure}

\begin{figure}[!t]
\centering
\includegraphics[width=0.99\columnwidth]{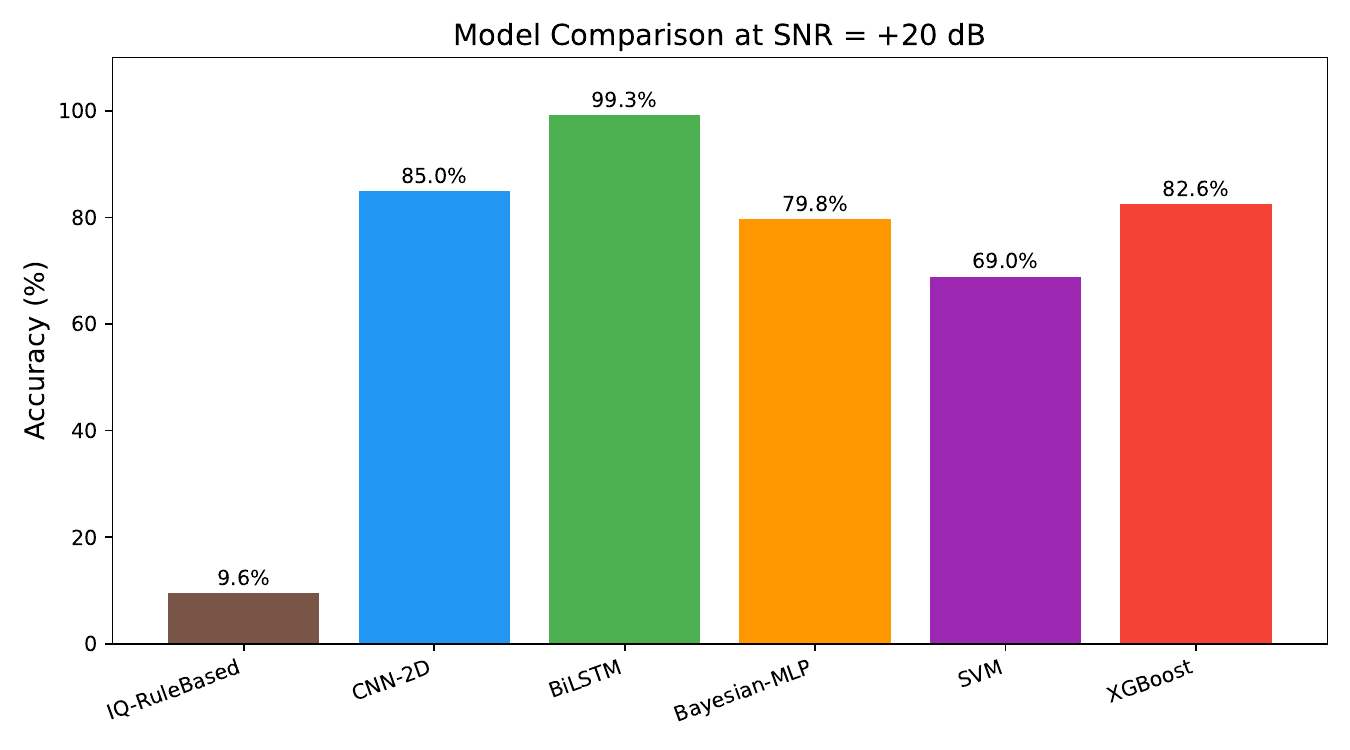}
\caption{Model accuracy at $+20$\,dB (same benchmark as Fig.~\ref{fig:snr}).}
\label{fig:bar}
\end{figure}

The 2D CNN, BiLSTM, Bayesian MLP, SVM, XGBoost, and a rule-based IQ analyzer were compared across seven SNR levels ($-5$ to $+25$\,dB) on a companion 199{,}920-sample dataset (2040 samples/class/SNR level, same taxonomy and methodology as Section~II-A, 70\%/10\%/20\% split, five runs averaged per model), conducted as part of the same research program prior to this manuscript and reused here rather than re-run. Fig.~\ref{fig:snr} presents the accuracy curves by SNR, and Fig.~\ref{fig:bar} the snapshot at $+20$\,dB: the 2D CNN exhibits the strongest low-SNR performance among fast methods (64.7\% at $-5$\,dB), the rule-based analyzer remains in the 6--15\% band, classical methods (SVM, XGBoost) show moderate performance, and BiLSTM reaches over 99\% at high SNR.

Table~\ref{tab:comparison} summarizes $+20$\,dB accuracy and inference time. The 2D CNN is the fastest at 0.138\,ms, giving the best accuracy-latency trade-off for the primary path; the BiLSTM, despite the highest accuracy, runs at 15.2\,ms and is engaged only when uncertainty triggers it.

\begin{table}[!h]
\caption{Method Comparison ($+20$\,dB, Companion Benchmark)}
\label{tab:comparison}
\centering
\small
\begin{tabular}{lrr}
\toprule
\textbf{Method} & \textbf{Accuracy} & \textbf{Latency} \\
\midrule
IQ Rule-Based              & 9.6\%           & 2.1\,ms \\
SVM                        & 69.0\%          & 1.0\,ms \\
Bayesian MLP               & 79.8\%          & 0.7\,ms \\
XGBoost                    & 82.6\%          & 0.8\,ms \\
\textbf{2D CNN (Proposed)} & \textbf{85.0\%} & \textbf{0.138\,ms} \\
BiLSTM                     & 99.3\%          & 15.2\,ms \\
\bottomrule
\end{tabular}
\end{table}

\subsection{Classification Analysis}

Class-wise analysis reveals a failure specific to the compact spectral feature space: the Bayesian MLP yields only 3\% recall for 4-FSK, which overlaps heavily with 2-tone FSK there and lacks temporal modeling. The 2D CNN partially captures symbol-timing via convolution (71\% recall), while the BiLSTM, using the full temporal IQ structure, resolves it almost without error.

This justifies the layered design: the uncertainty estimator produces a high $\mathcal{U}(x)$ for spectrally inseparable classes such as 4-FSK, delegating to the temporally-aware BiLSTM. Evaluation sets without 4-FSK~\cite{liu2017} hide this difficulty; our finding quantifies the limitation of compact spectral representations in FSK order disambiguation.

\subsection{MC Dropout Ablation Analysis}

\begin{figure}[!t]
\centering
\includegraphics[width=0.99\columnwidth]{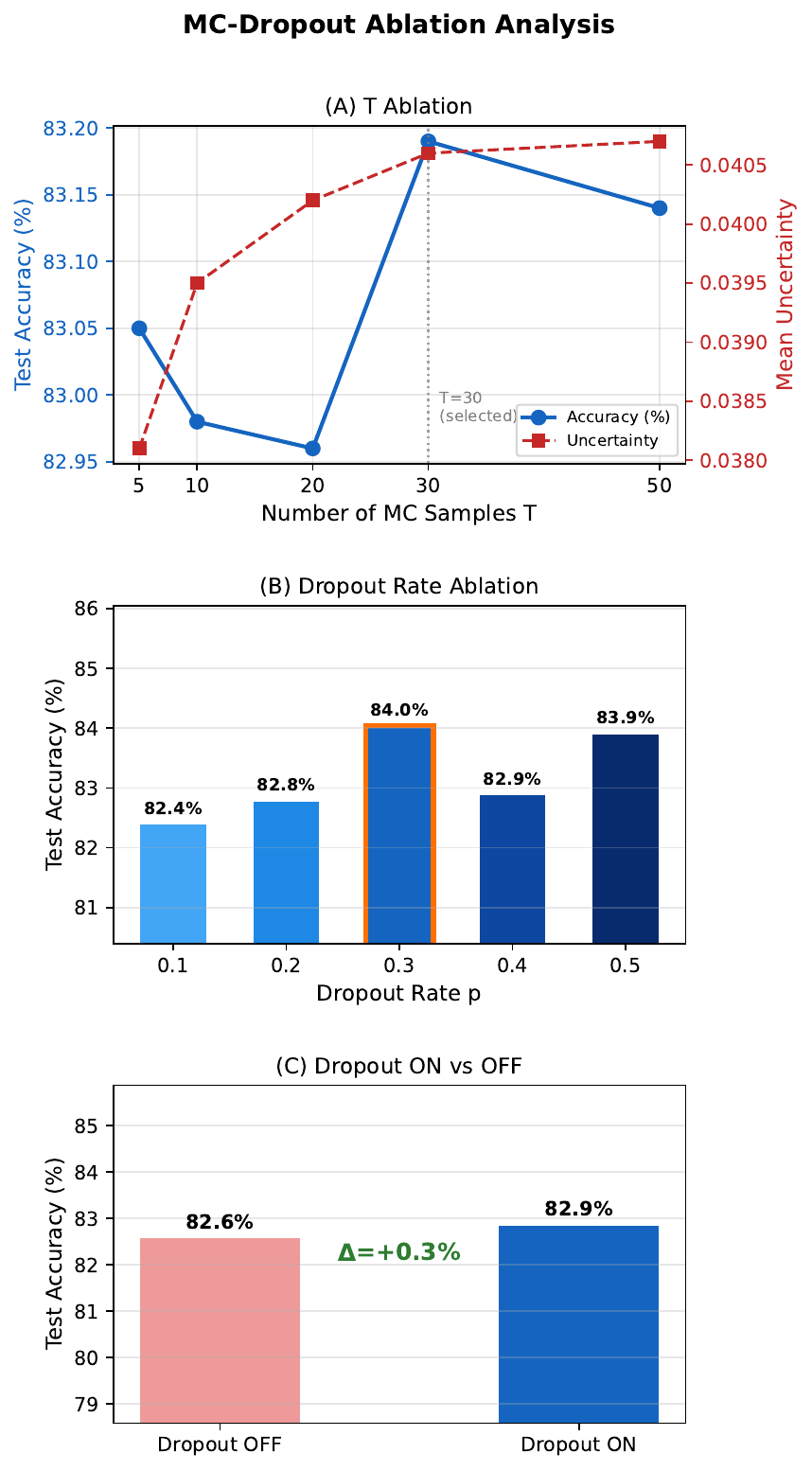}
\caption{MC Dropout ablation. (A) MC samples $T$: $T=30$ vs.\ $T=50$ differ by $<$0.2 points. (B) Dropout rate $p$: best at $p=0.3$. (C) With/without dropout: $+0.3$ points and Bayesian uncertainty capability.}
\label{fig:dropout}
\end{figure}

The uncertainty mechanism's design choices were examined with three ablations (Fig.~\ref{fig:dropout}). Varying MC samples $T\in\{5,10,20,30,50\}$ changed accuracy by only 0.1 points ($83.05\%\rightarrow83.14\%$) while inference cost grew linearly, confirming $T=30$ as a suitable balance. Dropout rates $p\in\{0.1,\dots,0.5\}$ peaked at $p=0.3$ (84.05\%), staying in the 82.3--84.1\% band, showing robustness. Finally, no-dropout reached 82.59\% versus 82.87\% with dropout ($p=0.4$); the small gap is expected, since dropout also enables Bayesian uncertainty estimation.

\subsection{End-to-End Hybrid Validation}
\label{sec:e2e}

The proposed hybrid algorithm (Algorithm~\ref{alg:hybrid}) was additionally validated end-to-end on an independently generated 29{,}400-sample dataset. Results confirm that the Bayesian MLP uncertainty score reliably predicts CNN misclassifications (AUROC 0.78 at $\tau=0.05$), with overall system performance summarized in Table~\ref{tab:e2e_summary}.

\begin{table}[!h]
\caption{End-to-End Hybrid Validation --- Summary}
\label{tab:e2e_summary}
\centering
\scriptsize
\setlength{\tabcolsep}{3.5pt}
\begin{tabular}{@{}ll@{}}
\toprule
\textbf{Parameter} & \textbf{Value} \\
\midrule
Dataset          & 29{,}400 samp.\ (300/class/SNR), 70/15/15 \\
Threshold $\tau$  & 0.05 (macro-F1-selected) \\
Unc.\ AUROC       & 0.78 \\
Hybrid acc.\      & 92.6\% (F1 92.3\%), 38.0\% routed \\
\bottomrule
\end{tabular}
\end{table}

\subsection{Channel Robustness Analysis}

\begin{figure}[!t]
\centering
\includegraphics[width=0.99\columnwidth]{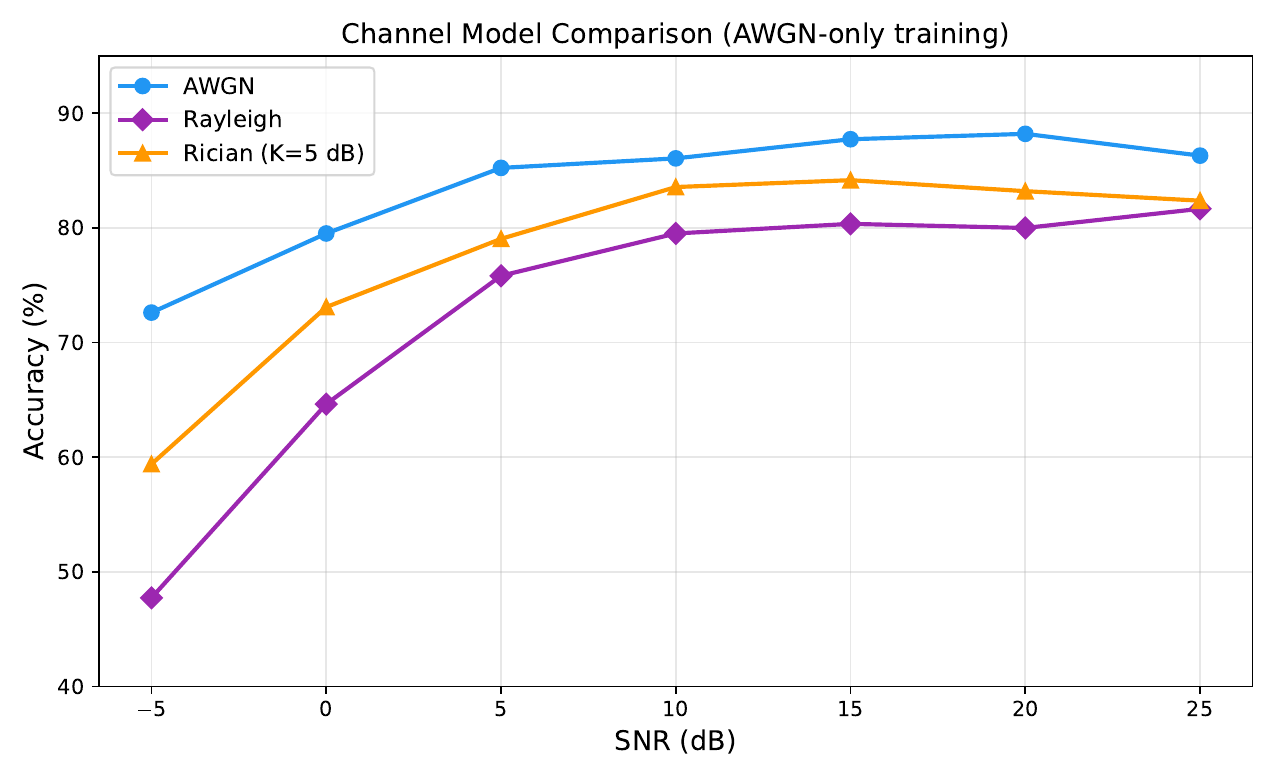}
\caption{Channel model comparison: AWGN, Rayleigh, Rician ($K=5$\,dB); model trained on AWGN only.}
\label{fig:channel}
\end{figure}

The AWGN-trained 2D CNN was tested on unseen Rayleigh ($h \sim \mathcal{CN}(0,1)$) and Rician ($K=5$\,dB) fading channels (Fig.~\ref{fig:channel}), attaining 88.2\% (AWGN), 83.2\% (Rician), and 80.0\% (Rayleigh) at $+20$\,dB -- Rayleigh drops to 47.7\% at $-5$\,dB, while Rician's line-of-sight component keeps it closer to AWGN -- quantifying the performance loss of an AWGN-only-trained model under fading.

\subsection{Channel-Adaptive Training}

To mitigate this loss, three training scenarios were compared: (i) AWGN only, (ii) Rayleigh only, and (iii) mixed (AWGN+Rayleigh+Rician) training. Table~\ref{tab:channel_train} summarizes the $+20$\,dB test results.

\begin{table}[!h]
\caption{Channel-Adaptive Training Comparison ($+20$\,dB Test Accuracy)}
\label{tab:channel_train}
\centering
\small
\begin{tabular}{lccc}
\toprule
\textbf{Training Channel} & \textbf{AWGN} & \textbf{Rayleigh} & \textbf{Rician} \\
\midrule
AWGN only          & \textbf{87.4\%} & 80.0\% & 82.9\% \\
Rayleigh only      & 85.6\% & 84.6\% & 85.7\% \\
Mixed (3 channels) & \textbf{88.2\%} & \textbf{85.6\%} & \textbf{86.5\%} \\
\bottomrule
\end{tabular}
\end{table}

Mixed training preserves AWGN accuracy while raising Rayleigh by 5.6 points and Rician by 3.6 points, showing it is an effective, low-cost adaptation strategy for fading channels.

\section{Conclusion}

In this study, an uncertainty-driven, multi-stage hybrid deep learning architecture was proposed for automatic RF signal recognition over a broad modulation space. The 2D CNN primary path provides 83.3$\pm$0.7\% accuracy at 0.138\,ms, clearly surpassing rule-based and classical ML methods, while the BiLSTM secondary path delivers over 99\% accuracy at high SNR; a preliminary end-to-end validation with a genuinely separate Bayesian MLP (Section~\ref{sec:e2e}) confirmed its uncertainty score predicts CNN misclassifications (AUROC 0.78) and raises hybrid accuracy to 92.6\% (macro-F1 92.3\%), though comprehensive OOD-rejection validation is left to future work.

A prominent finding is the Bayesian MLP's 3\% recall for 4-FSK, demonstrating the limitation of compact spectral representations in FSK order disambiguation and justifying the layered design. Ablations confirmed $T=30$ and $p=0.3$ as suitable, and mixed AWGN+Rayleigh+Rician training improved Rayleigh by 5.6 and Rician by 3.6 points without sacrificing AWGN performance.

All results were obtained in a controlled simulation environment. Future work includes validation on real SDR hardware, phase-preserving representations to resolve QPSK/8PSK confusion, and comparison with attention-mechanism architectures~\cite{wang2023ctgnet}.


\end{document}